\documentclass{appolb}
\usepackage{graphicx}
\usepackage{cite}
\usepackage{hyperref}
\usepackage{amsmath}
\usepackage{graphicx}
\usepackage{dcolumn}
\usepackage{bm}
\usepackage{graphicx}

\begin{document}
\title{Pseudorapidity and initial energy densities\\ 
in $p$$+$$p$ and heavy ion  collisions at RHIC and LHC%
\thanks{Presented by Ze-Fang Jiang at WPCF2018 in  Cracow, Poland, May 22-26, 2018}
}
\author{\underline{Ze-Fang Jiang}$^{~1,2}$, M. Csan\'ad$^{~3}$, G. Kasza$^{4}$ \\
 C. B. Yang$^{~1,2}$ and T. Cs\"org\H{o}$^{~4,5}$
\address{{$^1$ Key Lab. of Quark and Lepton Physics,  Wuhan, 430079, China}\\
{$^2$ Inst. of Particle Physics, CCNU, Wuhan, 430079, China}\\
{$^3$ ELTE,  H-1117 Budapest, P{\'a}zm{\'a}ny P. s. 1/A, Hungary}\\
{$^4$ EKU KRC, H-3200 Gy{\"o}ngy{\"o}s, M{\'a}trai \'ut 36, Hungary}\\
{$^5$ Wigner RCP,  H-1525 Budapest 114, P.O.Box 49, Hungary}
}
}
\maketitle
\begin{abstract}
	A known exact and accelerating solution of  relativistic hydrodynamics
	for perfect fluids is utilized to describe 
	pseudorapidity densities of $\sqrt{s_{NN}} =5.02$ TeV Pb$+$Pb and
	$\sqrt{s}=13$ TeV $p$$+$$p$ collisions at LHC. 
	We evaluate a conjectured initial energy densities
	$\epsilon_{\rm corr}$ 
	in these collisions, 
	and compare them to Bjorken's initial energy density estimates,  and to
	results for Pb$+$Pb collisions at
	$\sqrt{s_{NN}} = 2.76$ TeV and $p$$+$$p$ collisions at $\sqrt{s} = 7$ and
	$8$ TeV. 
\end{abstract}

\section{Introduction}
Relativistic hydrodynamics is an efficient theoretical framework to study the
properties of strongly interacting Quark Gluon Plasma (sQGP), produced in relativistic heavy ion
collisions~\cite{Bass1998vz,Gyulassy:2004zy}.  Both analytical 
and numerical results of hydrodynamics highlighted important  details of
the time evolution of sQGP~\cite{Landau:1953gs,Hwa:1974gn,Bjorken:1982qr,
Csorgo:1995pr,Biro:2000nj,Csorgo:2003rt,Csorgo:2006ax,Csorgo:2008prc,
Nagy2009eq,Gubser:2010ze} as reviewed in Ref.\@~\cite{deSouza:2015ena}. A brief review on
the successful applications of exact analytic solutions of relativistic hydrodynamics to describe
the evolution of longitudinal phase-space in high energy collisions was given recently in Section 2 of
Ref.\@~\cite{Csorgo:2018pxh}.  

In recent publications~\cite{Csorgo:2018pxh,Csand:2016arx,Ze-Fang:2017ppe}, the pseudorapidity
distributions of various colliding systems were analyzed to study the
longitudinal expansion dynamics at RHIC and LHC energies.  These works were
based on an accelerating and  explicit, but rather academic family of exact solutions of
relativistic hydrodynamics, as found by Cs\"org\H{o},~Nagy, and~Csan\'ad (CNC)~
in Refs.\@~\cite{Csorgo:2006ax,Csorgo:2008prc}. Given that the selected CNC solutions
were 1+1 dimensional, the transverse momentum
distributions were phenomenologically modelled utilizing also the 1+3 dimensional Buda-Lund hydro
model~\cite{Csorgo:1995pr}.  The initial thermodynamic quantities for
$\sqrt{s_{NN}}=200$ GeV Cu$+$Cu, $\sqrt{s_{NN}}=130$  and 200 GeV Au$+$Au,
$\sqrt{s_{NN}}=2.76$ TeV Pb$+$Pb, and $\sqrt{s}=7$ and 8 TeV $p$$+$$p$ collisions
at RHIC and LHC energies were estimated and published recently in Refs.\@~\cite{Csand:2016arx,Ze-Fang:2017ppe},
so we do not detail them here, due to space limitations.  Instead, we present new results for the
pseudorapidity distributions and for the initial energy densities at top LHC energies, 
for Pb$+$Pb collisions at $\sqrt{s_{NN}}= 5.02$ TeV and $p$$+$$p$ collisions 
at $\sqrt{s}=13$ TeV. 

\section{Accelerating hydrodynamics and initial energy densities }
The dynamical equations of relativistic perfect fluid hydrodynamics 
correspond to the local conservation entropy and four-momentum:
\begin{eqnarray}
	\partial_{\mu} (\sigma u^{\mu}) & = & 0,\\
	\partial_{\nu} T^{\mu\nu}& =& 0, 
\label{Tmn}
\end{eqnarray}
where 
the entropy density is denoted by $\sigma$, 
the four velocity field by $u^{\mu}$, 
the energy density by $\epsilon$,
the  pressure by $p$ and the energy-momentum four-tensor of perfect fluids is
$T^{\mu\nu}=(\epsilon+p)u^{\mu}u^{\nu}-pg^{\mu\nu}$.
The equation of state $\epsilon=\kappa p$ closes the  above set of dynamical equations.
For the case of vanishing baryochemical potential $\mu_B = 0$, the fundamental thermodynamical
relation $\epsilon + p = T \sigma$ can also be utilized to solve these equations, 
and here we also assume that $\kappa = 1/c_s^2 \ne \kappa(T)$, so the speed of sound $c_s$ is modelled with
a temperature $T$ independent, average value. An accelerating but rather academic family of exact solutions was
detailed in Refs.\@~\cite{Csorgo:2006ax,Csorgo:2008prc}:
\begin{eqnarray}
	u^{\mu} & = &(\cosh(\lambda\eta_{x}),\sinh(\lambda\eta_{x})),  \\
	p 	& = & p_{0}\left(\frac{\tau_{0}}{\tau}\right)^{\lambda \frac{\kappa+1}{\kappa}},
\label{p1}
\end{eqnarray}
where the longitudinal proper time is denoted by $\tau=\sqrt{t^{2}-r_z^{2}}$, 
the space-time rapidity is denoted by $\eta_{x}=0.5\log\left[(t+r_z)/(t-r_z)\right]$, and here
we limit the discussion only  to 1+ 1 dimensional solutions with $x^{\mu} =( t,
r_z)$ and $u^{\mu} = (u^0,u_z)$, that corresponds to one of the five different classes
of solutions that were detailed in Refs.\@~\cite{Csorgo:2006ax,Csorgo:2008prc}.  
In these solutions, the longitudinal acceleration parameter is  a free fit parameter, denoted by  $\lambda$ and
the initial values for the pressure and thermalization time are denoted by
$p_{0}$ and $\tau_{0}$, respectively. The price for the freedom in $\lambda$ was
a fixed value of super-hard equation of state, $\kappa = 1$. 
Combining the exact solution of
relativistic hydrodynamics with a Cooper-Frye flux term, and embedding this
solution to 1 + 3 dimensions with $x^{\mu } = (t, r_x, r_y, r_z)$ 
and $p^{\mu} = (E, p_x, p_y, p_z)$, 
the rapidity distribution $dn/dy$ was obtained in a saddle-point approximation as 
~\cite{Csorgo:2006ax,Csorgo:2008prc}
\begin{equation}
\frac{dn}{dy} = 
	\left.\frac{dn}{dy}\right\vert_{y= 0} 
	\cosh^{-\frac{\alpha}{2}-1}\left(\frac{y}{\alpha}\right)\exp\left\{-\frac{m}{T_{f}}
	    \left[\cosh^{\alpha}\left(\frac{y}{\alpha}\right) - 1 \right]\right\},
\label{dndy}
\end{equation}
where $\alpha=\frac{2\lambda-1}{\lambda-1}$, 
the freeze-out temperature is denoted by $T_f$, the mass of particles is $m$ and 
the rapidity of the observed particles is denoted by $y=0.5\log((E+p_z)/(E-p_z))$.
The constant of normalization $ \left.\frac{dn}{dy}\right\vert_{y= 0} $ is proportional to 
$S_{\perp}$, that  stands for the transverse cross section of the fluid.

The pseudorapidity density distribution $\frac{dn}{d\eta}$, with the help of an advanced saddle-point integration is
given~\cite{Csorgo:1995pr,Csorgo:2006ax} as a parametric curve $(\eta(y),
\frac{dn}{d\eta}(y) )$, where the parameter is the rapidity $y$:
\begin{equation}
	\left(\eta(y) \, , \frac{dn}{d\eta}(y)  \right) =
	\left( \frac{1}{2}\log\left[\frac{\bar{p}(y) + \bar{p}_z(y)}{\bar{p}(y)-\bar{p}_z(y)}\right] \, , 
	 \frac{\bar{p}(y)}{\strut \bar{E}(y)}\frac{dn}{dy} \right),
\label{8}
\end{equation}
where $\bar{A}(y)$ denotes  the rapidity dependent average value of the variable $A$ including
the various components of the four-momentum, and
the Jacobian connecting the double differential ($y$, $m_t$)  and ($\eta$, $m_t$)
distributions has been utilized at the average value of the transverse momentum~\cite{Csorgo:2006ax}. 
Based on the Buda-Lund hydrodynamic
model~\cite{Csorgo:1995pr}, in the region of $p_{T}$ $<$ 2 GeV, the relation
between mean transverse momentum $\bar{p}_{T}$ and the the effective
temperature $T_{\rm eff}$ at a given rapidity $y$ can be written as
\begin{eqnarray}
	\bar{p}_{T}(y) & = & \frac{T_{\rm eff}}{1+\frac{\sigma_T^{2}}{2}(y-y_{\rm mid})^{2}}, \label{9}
\end{eqnarray}
where $\sigma_T$ parameterizes the rapidity dependence of the average transverse momentum. 
In our case,  $\sigma_T$ and $T_{\rm eff}$  are free fit parameters. Their values can be determined
either from fits to data on the rapidity dependent transverse momentum spectra,  or phenomenologically as in
Ref.\@~\cite{Csorgo:1995pr} or dynamically, as in Ref.\@~\cite{Csorgo:2018pxh}.
Midrapidity is denoted by $y_{\rm mid}$. In our case, it is at $y_{\rm mid} = 0$.

Our fit results to pseudorapidity densities allow for advanced estimates of the initial energy densities.  
The Bjorken-estimate~\cite{Bjorken:1982qr} at midrapidity is
\begin{equation}
\epsilon_{\rm Bj}=\frac{1}{S_{\perp}\tau_{0}}\frac{dE_{T}}{d\eta}
=\frac{\langle E_{T}\rangle}{S_{\perp}\tau_{0}}\frac{dn}{dy}.
\label{11}
\end{equation}
In case of a longitudinally accelerating flow, 
the acceleration effects modify Bjorken's estimate.
A conjectured initial energy density 
$\epsilon_{\rm corr}$ ~\cite{Csand:2016arx,Ze-Fang:2017ppe}
that corrects Bjorken's estimate for acceleration effects reads as
\begin{equation}
\epsilon_{\rm corr} = (2\lambda-1)\left(\frac{\tau_{f}}{\tau_{0}}\right)^{\lambda-1}\left(\frac{\tau_{f}}{\tau_{0}}\right)^{(\lambda-1)(1-\frac{1}{\kappa})}\epsilon_{\rm Bj}.
\label{12}
\end{equation}
This estimate explicitely takes into account the bending of the fluid world-lines due to acceleration.
However, it is based on results that are obtained exactly in the $\kappa=1$ case only.
Until most recently, the dependence of the initial energy density on the speed of sound $c_s = 1/\sqrt{\kappa} $ 
had not been derived exactly, only a conjecture was known so far. Given that the speed of sound is 
is an important physical property of the sQGP,
it is important to cross-check and derive exact results for realistic values of the speed of sound,
corresponding to $c_s^2 \approx 0.1$. 
However, let us emphasize that this conjecture, Eq.~(\ref{12}) is based on the determination of the  acceleration parameter $\lambda$ 
from fits to the measured  pseudorapidity density distributions. The dependence of the initial energy density on the
initial and  freeze-out proper-times,  $\tau_{0}$ and $\tau_{f}$, is a topic of ongoing research, with first results presented in
Refs.\@~\cite{Csorgo:wpcf2018,Kasza:wpcf2018}.

\section{Results}

\begin{figure}
\vspace{-10pt}
\begin{center}
\includegraphics[width=0.47\linewidth]{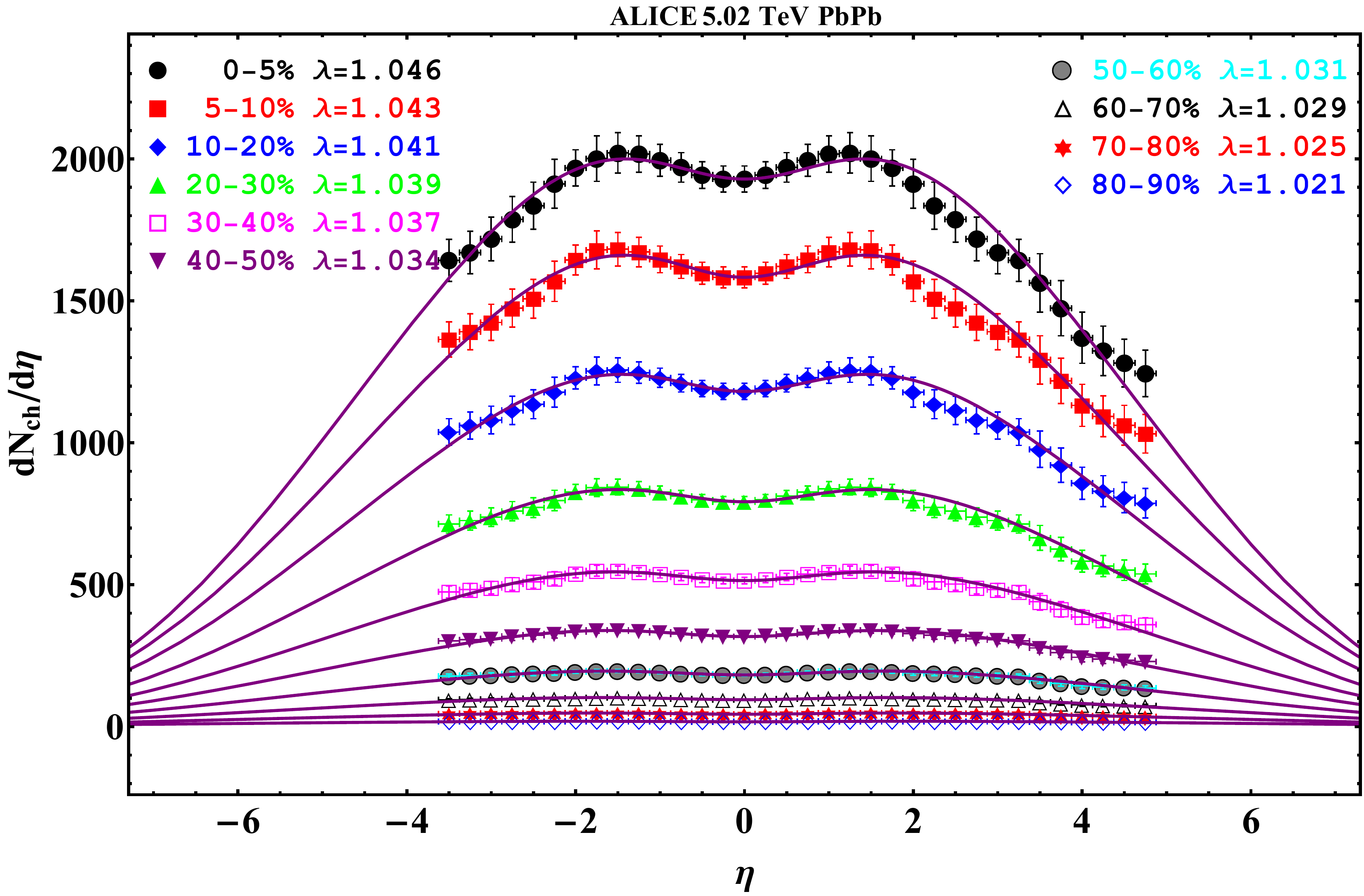}
\includegraphics[width=0.47\linewidth]{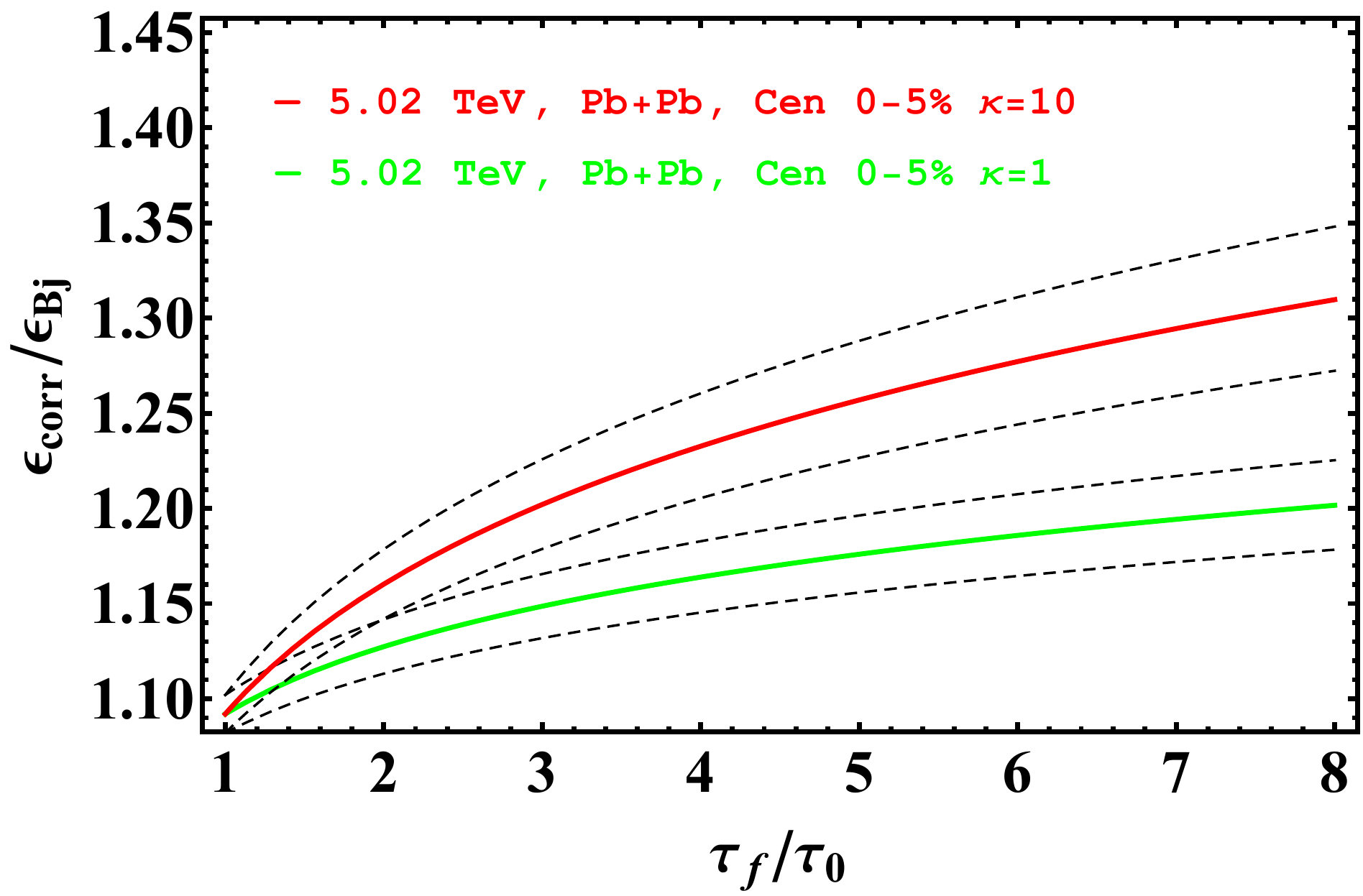}
\end{center}
	\vspace{-10pt}
	\caption{(Color online) The left panel shows hydrodynamical fits using Eqs.~ (\ref{dndy}-\ref{9}) to 
	$dn_{\rm ch}/d\eta$ data  as measured by the  ALICE Collaboration in $\sqrt{s_{NN}} = 5.02$ TeV Pb$+$Pb collisions. 
	The right panel indicates with solid curves the $\epsilon_{\rm corr}/\epsilon_{\rm Bj}$ correction factor,
	as a function of the ratio of freeze-out time and thermalization time $\tau_{f}/\tau_{0}$,
	for the centrality class of 0-5 \%, both for the exact solution with $\kappa = 1$ superhard equation of state,
	and for the conjectured energy density values for the realistic $\kappa = 10$ soft equation of state,
	while the dashed lines represent the uncertainty of these estimates as determined from the errors of the fit parameters.
	}
\label{f:pbpbfits}
\end{figure}

\begin{figure}
\vspace{-10pt}
\begin{center}
\includegraphics[width=0.47\linewidth]{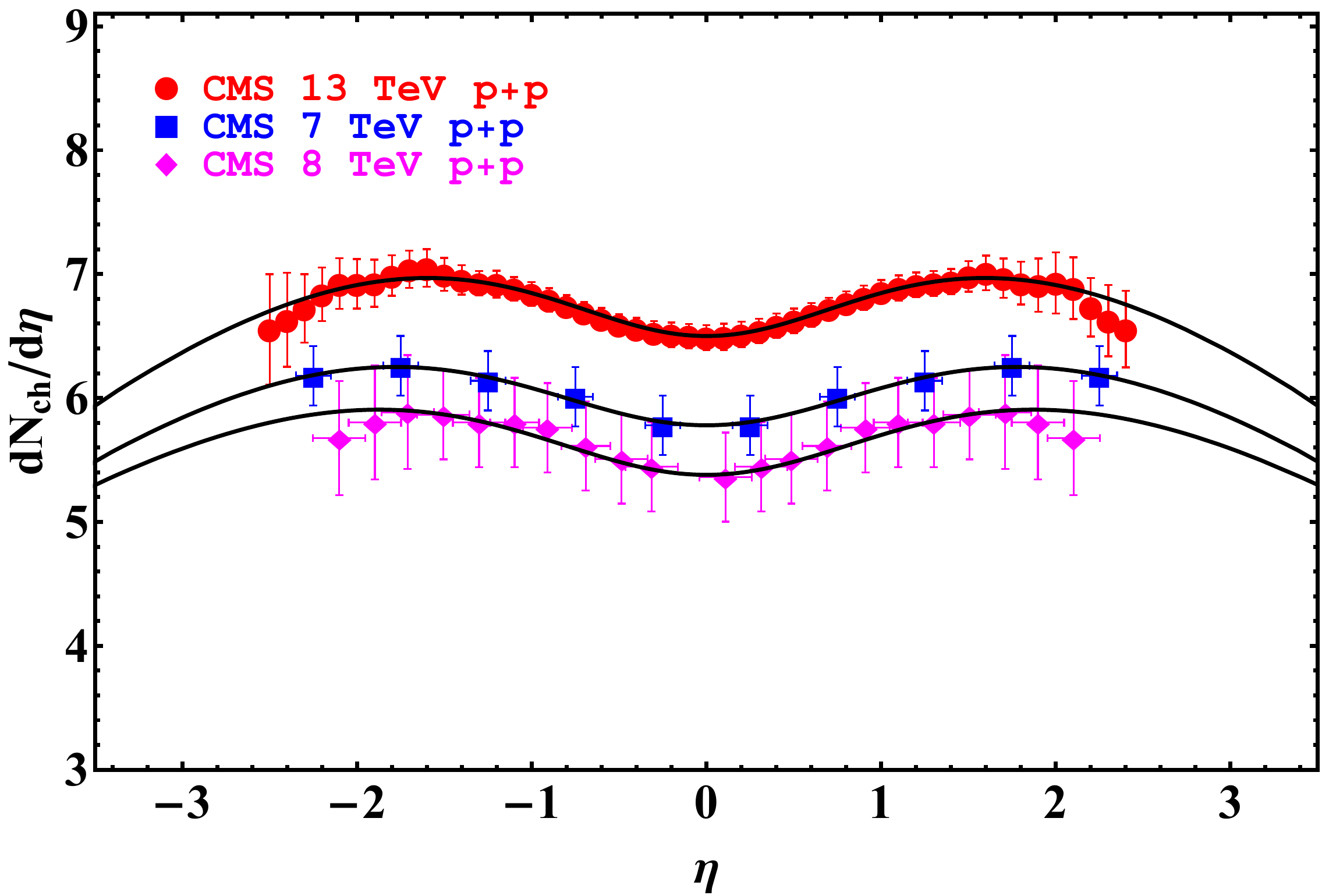}
\includegraphics[width=0.47\linewidth]{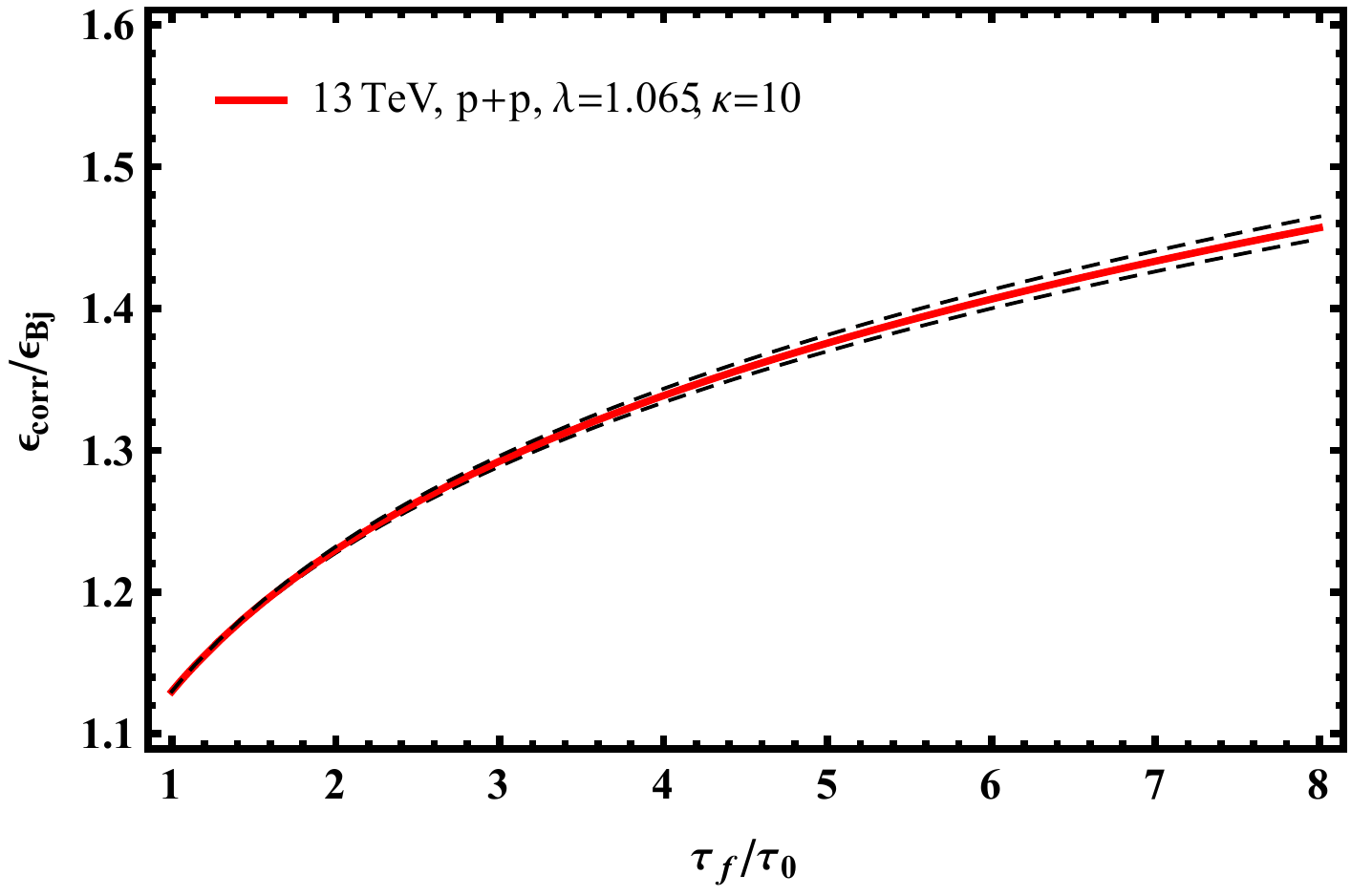}
\end{center}
	\vspace{-10pt}
	\caption{
		Same as Fig.~\ref{f:pbpbfits}, but for $p$$+$$p$ collisions at $\sqrt{s}=$$13$ TeV.
	}
\label{f:pp}
\end{figure}
\begin{table}
\begin{center}
\begin{tabular}{l c c c c }
\hline\hline
$\sqrt{s}$ & ~$\left.\frac{dn}{d\eta}\right\vert_{\eta=\eta_{0}}$ &~$\lambda$  &~$~\epsilon_{\rm Bj} [{\rm GeV/fm^{3}}]$  &~$~\epsilon_{\rm corr} [{\rm GeV/fm^{3}}]$    \\  [0.3 ex]
\hline
~$2.76$ TeV    &~1615$\pm$39      &~1.050$\pm$0.005        &~12.50$\pm$0.44    &~15.07$\pm$0.81      \\
~$5.02$ TeV   & ~1929$\pm$46      &~1.046$\pm$0.013        &~14.85$\pm$0.53    &~17.40$\pm$0.61    \\
\hline\hline
\end{tabular}
\end{center}
\caption{Acceleration parameters and initial energy density estimations for 2.76~\cite{Ze-Fang:2017ppe} and 5.02 TeV 0-5\% centrality Pb$+$Pb data~\cite{Adam:2016ddh}. Auxiliary values of $T_{f}= 90$ MeV, $T_{\rm eff}$=0.27$\pm$0.03 GeV, $\bar{m}=0.24$ GeV, $\sigma_T=0.9\pm0.1$ have been utilized, based on Refs.\@~\cite{Csorgo:1995pr,Csand:2016arx,Ze-Fang:2017ppe}.}\label{t:pbpb1}
\end{table}

\begin{table}
\begin{center}
\begin{tabular}{l c c c c }
\hline\hline
$\sqrt{s}$ & ~$\left.\frac{dn}{d\eta}\right\vert_{\eta=\eta_{0}}$ &~$\lambda$  &~$~\epsilon_{\rm Bj} [{\rm GeV/fm^{3}}]$  &~$~\epsilon_{\rm corr} [{\rm GeV/fm^{3}}]$  \\ [0.3 ex]
\hline
~$7$ TeV    &~5.78$\pm$0.01      &~1.073$\pm$0.001        &~0.51$\pm$0.01     &~ 0.64$\pm$0.01     \\
~$8$ TeV    &~5.36$\pm$0.02      &~1.067$\pm$0.001        &~0.52$\pm$0.01    &~0.64$\pm$0.01      \\
~$13$ TeV   &~6.50$\pm$0.02      &~1.065$\pm$0.013        &~0.56$\pm$0.02    &~0.69$\pm$0.02    \\
\hline\hline
\end{tabular}
\end{center}
	\caption{Acceleration parameters and initial energy density estimations for $\sqrt{s} =$ $7$, $8$~\cite{Csand:2016arx} and $13$ TeV $p$$+$$p$ data~\cite{Sirunyan:2017zmn}. Auxiliary values of $T_{f}=T_{\rm eff}=0.17\pm0.01$ GeV, $\bar{m}=0.14$ GeV, $\sigma_T=0.81\pm0.04$ have been utilized, based on Refs.\@~\cite{Csorgo:1995pr,Csand:2016arx,Ze-Fang:2017ppe}.}\label{t:pp1}
\end{table}

Measurements of the charged particle pseudorapidity distribution
$dn/d\eta$ for $\sqrt{s_{NN}}=$ 5.02 TeV Pb$+$Pb collisions and $\sqrt{s}=$ 13
TeV $p+p$ collisions were presented by the ALICE ~\cite{Adam:2016ddh}
and the CMS Collaborations~\cite{Sirunyan:2017zmn}. Here we extract the
acceleration parameter $\lambda $ of these collisions and apply it to calculate the
energy density correction ratio $\epsilon_{\rm corr}/\epsilon_{\rm Bj}$ as a
function of $\tau_{f}/\tau_{0}$). Fit results to ALICE and CMS
data are shown in Figs.~\ref{f:pbpbfits} and \ref{f:pp}. Our advanced 
estimates of the initial energy densities  $\epsilon_{\rm
corr}$ are given   in Tables~\ref{t:pbpb1} and \ref{t:pp1}, for the squared speed of
sound $c_{s}^{2} = 0.1$ and $\tau_{f}/\tau_{0}=6\pm2$.

\section{Summary and conclusions}
	We have evaluated the conjectured initial energy densities
	$\epsilon_{\rm corr}$ in $p$$+$$p$ and in Pb$+$Pb collisions at the currently
	available highest LHC energies,  and compared them to Bjorken's initial energy density estimates as well as to
	earlier results for Pb$+$Pb collisions at
	$\sqrt{s_{NN}} = 2.76$ TeV and $p$$+$$p$ collisions at $\sqrt{s} = 7$ and
	$8$ TeV. 
	Our new results are similar to our recent results published in
	Ref.\@~\cite{Csorgo:2018pxh}. 
	Our results were found to be not inconsistent --
	neither in  proton-proton  nor in heavy ion reactions -- with longitudinal
	expansion dynamics of hydrodynamical origin.



\section*{Acknowledgements}
We thank M. Kucharczyk, Mariola Kłusek-Gawenda and 
the Organizers of WPCF 2018 for their kind hospitality and for an inspiring and useful meeting.
Our research has been partially supported by the
bilateral Chinese-Hungarian intergovernmental grant No.~T\'ET 12CN-1-2012-0016,
the CCNU PhD Fund 2016YBZZ100 of China,
the COST Action CA15213, THOR Project of the European Union,
the Hungarian NKIFH grants No. FK-123842 and FK-123959,
the Hungarian EFOP 3.6.1-16-2016-00001 project,
the NNSF of China under grant No.~11435004
and by the exchange programme of the Hungarian and the Ukrainian Academies of Sciences, grants
NKM-82/2016 and NKM-92/2017.
M. Csan\'ad was partially supported by the J\'anos Bolyai Research Scholarship and the \'UNKP-17-4 New National Excellence
Program of the Hungarian Ministry of Human Capacities.


\begin{thebibliography}{90}
\bibitem{Bass1998vz}
 S. A. Bass, M. Gyulassy, H. St\"oecker, W. Greiner,
 J. Phys. G{\bf 25},~R1 (1999).

\bibitem{Gyulassy:2004zy}
 M.~Gyulassy, and L.~McLerran,
 Nucl. Phys. A {\bf 750}, 30-63 (2004).

 \bibitem{Landau:1953gs}
  L.~D.~Landau,
  Izv.\ Akad.\ Nauk Ser.\ Fiz.\ {\bf 17}, 51 (1953).

\bibitem{Hwa:1974gn}
  R.~C.~Hwa,
 Phys.\ Rev.\ D{\bf 10}, 2260 (1974).

\bibitem{Bjorken:1982qr}
  J.~D.~Bjorken,
 Phys.\ Rev.\ D {\bf 27}, 140 (1983).

\bibitem{Biro:2000nj}
 T.~S.~Bir{\'o},
 Phys. Lett. B{\bf 487}, 133 (2000).

\bibitem{Csorgo:1995pr}
 T.~Cs\"org\H{o} and B. L\"orstad,
 Phys.\ Rev.\ C{\bf54}, 1390(1996).

\bibitem{Csorgo:2003rt}
 T.~Cs\"org\H{o}, F.~Grassi, Y.~Hama, T.~Kodama,
  Phys. Lett.~B{\bf 565}, 107 (2003).

\bibitem{Csorgo:2006ax}
 T.~Cs\"org\H{o}, M. I.~Nagy, and M.~Csan\'ad,
 Phys. Lett. B{\bf 663}, 306 (2008).

\bibitem{Csorgo:2008prc}
 M. I. Nagy, T.~Cs\"org\H{o}, and M.~Csan\'ad,
 Phys.\ Rev.\ C{\bf77}, 024908(2008).

\bibitem{Nagy2009eq}
 M. I.~Nagy,
  Phys. Rev.~C {\bf 83}, 054901 (2011).

\bibitem{Gubser:2010ze}
 S.~S.~Gubser,
 Phys. Rev.~D {\bf 82}, 085027 (2010).


\bibitem{deSouza:2015ena} 
  R.~Derradi de Souza, T.~Koide and T.~Kodama,
  Prog.\ Part.\ Nucl.\ Phys.\  {\bf 86}, 35 (2016)

\bibitem{Csorgo:2018pxh} 
	T.~Cs\"{o}rg\H{o}, G.~Kasza, M.~Csan\'ad and Z.F.~Jiang,
  Universe {\bf 4}(6), 69 (2018). 


\bibitem{Csand:2016arx}
 M.~Csan\'ad, T.~Cs\"org\H{o}, Z. F. Jiang and C. B. Yang,
 Universe  {\bf 3} (1), 9 (2017).


\bibitem{Ze-Fang:2017ppe} 
	Z.F. Jiang, C.B. Yang, M.~Csan\'ad, T.~Cs\"org\H{o},
 Phys. Rev.~C{\bf 97}, 064906 (2018).

\bibitem{Adam:2016ddh}
 J. Adam and {\it et al.}  [ALICE Collaboration],
 Phys. Lett.~B{\bf 772}.~567.

\bibitem{Sirunyan:2017zmn}
 Albert M.~Sirunyan~{\it et al.}  [CMS, TOTEM Collaboration],
 Phys. Rev. D{\bf 96}. 112003 (2017).

\bibitem{Csorgo:wpcf2018}
	T. Cs\"org\H{o}, G. Kasza, M. Csan\'ad and Z.-F. Jiang, 
	Talk at WPCF 2018,
	\url{https://indico.ifj.edu.pl/event/199/contributions/1111/}

\bibitem{Kasza:wpcf2018}
	G. Kasza and T. Cs\"org\H{o},
	Talk at WPCF 2018,
	\url{https://indico.ifj.edu.pl/event/199/contributions/1106/}

\end{thebibliography}
\end{document}